\documentclass[aps,pre,floats,twocolumn,showpacs,superscriptaddress]{revtex4}

\usepackage{epsfig}
\usepackage{latexsym,amsmath}

\newcommand{\avk}{\langle k \rangle} \newcommand{\fluck}{\langle k^2 \rangle}
\newcommand{\avh}{\langle h \rangle} \newcommand{\fluch}{\langle h^2 \rangle}

\newcommand{\condPN}[2]{P({#1} \, \vert \, {#2})}
\newcommand{\condpN}[2]{p({#1} \, \vert \, {#2})}
\newcommand{\prop}[2]{g({#1} \, \vert \, {#2})}
\newcommand{\tprop}[2]{\hat{g}({#1} \, \vert \, {#2})}
\newcommand{\pprop}[3]{g^{(h)}_{{#1}}({#2} \, \vert \, {#3})}
\newcommand{\tpprop}[3]{\hat{g}^{(h)}_{{#1}}({#2} \, \vert \, {#3})}

\begin{document}

\title{Class of correlated random networks with hidden variables}

\author{Mari{\'a}n Bogu{\~n}{\'a}}

\affiliation{Departament de F{\'\i}sica Fonamental, Universitat de
  Barcelona, Av. Diagonal 647, 08028 Barcelona, Spain}

\author{Romualdo Pastor-Satorras}

\affiliation{Departament de F{\'\i}sica i Enginyeria Nuclear, Universitat
  Polit{\`e}cnica de Catalunya, Campus Nord, M{\'o}dul  B4, 08034 Barcelona, Spain}

\date{\today}

\begin{abstract}
  We study a class models of correlated random networks in which
  vertices are characterized by \textit{hidden variables} controlling
  the establishment of edges between pairs of vertices. We find
  analytical expressions for the main topological properties of these
  models as a function of the distribution of hidden variables and the
  probability of connecting vertices.  The expressions obtained are
  checked by means of numerical simulations in a particular example.
  The general model is extended to describe a practical algorithm to
  generate random networks with an \textit{a priori} specified
  correlation structure.  We also present an extension of the class,
  to map non-equilibrium growing networks to networks with hidden
  variables that represent the time at which each vertex was
  introduced in the system.
\end{abstract}


\maketitle

\section{Introduction}

A large effort has been recently devoted to the study of a very
large ensemble of interacting systems that can be described in
terms of complex networks (or graphs), in which the vertices
represent typical units and the edges represent the interactions
between pairs of units \cite{barabasi02,dorogorev,mendesbook}.
Stimulated by this finding, a theory of complex networks, deeply
rooted in the classical graph theory \cite{bollobas98}, has hence
been developed, finding fruitful applications in fields as diverse
as the Internet \cite{falou99,calda00,alexei,alexei02}, the
World-Wide-Web \cite{www99}, social communities \cite{strog01},
food-webs \cite{montoya02}, or biological interacting networks
\cite{wagner01,jeong01,spsk,vazquez}.

The study of complex networks, boosted by the new availability of
powerful computers capable to deal with very large databases, was
initially focused in the study of global properties, such as the
average shortest path length, the average clustering coefficient, or
the degree distribution \cite{barabasi02,dorogorev,mendesbook}.  This
work led to the discovery that most natural complex networks usually
exhibit two typical properties: (i) The \textit{small-world} property
\cite{watts98}, that is defined by an average path length---average
distance between any pair of vertices---increasing very slowly
(usually logarithmically) with the network size $N$. (ii) A
\textit{scale-free} degree distribution. If we define the degree
distribution, $P(k)$, as the probability that a vertex is connected to
$k$ other vertices, then scale-free networks are characterized by a
power-law behavior $P(k) \sim k^{-\gamma}$, where $\gamma$ is a characteristic
degree exponent. These properties imply a large connectivity
heterogeneity and a short average distance between vertices, which
have considerable impact on the behavior of physical processes taking
place on top of the network, such as the resilience to random damage
\cite{barabasi00,newman00,havlin01} or the spreading of infective
agents \cite{pv01a,pv01b,lloydsir,moreno02}.

It was soon realized, however, that these properties do not provide a
sufficient characterization of natural networks. In particular, these
systems seem to exhibit also ubiquitous degree \textit{correlations},
which translate in the fact that the degrees of the vertices at the
end points of any given edge are not independent
\cite{alexei,alexei02,assortative,newmanmixing}. This observation has
led to a classification of networks according to the nature of their
degree correlations \cite{assortative}: In the presence of positive
correlations (vertices with large degree tend to connect more
preferably with vertices with large degree), the network is said to
show \textit{assortative mixing}. On the other hand, negative
correlations (highly connected vertices are preferably connected to
vertices with low degree) imply the presence of \textit{dissortative
  mixing}.  At the same time, it has been pointed out that the
presence of correlations might have important consequences in
dynamical processes taking place in the topology defined by the
network \cite{marian1,morenostructured,morenopercolation,marian3}.
Motivated by these observations, several works have been recently
devoted to set up a general framework to study the origin of
correlations in random networks \cite{bergcorrels,mendescorrels}.
At this respect, it is particularly interesting the models
introduced by Caldarelli \textit{et al.} \cite{gcalda03} and
S{\"o}derberg \cite{soderberg}.  These models consider graphs in
which each vertex has assigned a tag (\textit{type} or
\textit{fitness}), randomly drawn from a fixed probability
distribution.  Edges are assigned to pairs of vertices with a
given connection probability, depending on the values of the tags
assigned at the edge end points. This construction generates
random networks which exhibit peculiar correlation and percolation
properties \cite{gcalda03,soderberg}.

In this paper we present a generalization of the models described in
Ref.~\cite{gcalda03,soderberg}, that can be encompassed in a general
class of models with \textit{hidden variables} tagging the vertices,
and completely determining the topological structure of the ensuing
network.  We develop a detailed analysis of the correlations present
in this class of network models, providing explicit analytical
expressions for both two and three vertices degree correlations. We
distinguish between sparse networks (with finite average degree
$\avk$) and non-sparse networks (with diverging $\avk$ for a number of
vertices $N\to\infty$).  Even though both cases are enclosed in this class
of networks, analytical expressions are simpler in the former case.
As an example of our formalism, we consider the intrinsic fitness
model introduced in Ref.~\cite{gcalda03}, which belongs to the subset
of non-sparse networks, and that has attracted a great deal of
attention as an alternative to generate scale-free networks without
growth nor preferential attachment \cite{barab99}. The solution of
this model in the continuous degree approximation is compared with
extensive numerical simulations, yielding a remarkable agreement for
all the topological properties considered. As a particular case of the
general class of models with hidden variables, we propose a practical
algorithm to generate correlated random networks with a given
correlation structure. The algorithm levers in the assignation of
hidden variables with the structure of the degrees of a real network.
Following this approach, it is possible to easily generate networks
matching any desired correlation pattern, as we show by means of
analytical calculations and numerical simulations. Finally, we present
the extension of this class of models to non-equilibrium growing
networks. By mapping the hidden variables to the time in which
vertices are introduced in the network \cite{soderberg}, and by means
of an appropriately chosen connection probability, we define an
algorithm that yields networks exhibiting all the properties (in
particular aging) exhibited by traditional scale-free growing models.

The paper is organized as follows. In Sec.~\ref{sec:meas-corr-compl}
we review some general results concerning the measure of correlations
in complex networks, which will be useful through the rest of the
paper. In Sec.~\ref{sec:hidd-vari-models} we introduce the general
analytical formulation of the class of correlated networks with hidden
variables. Sec.~\ref{sec:intr-fitn-model} is devoted to the analytical
and numerical study of the intrinsic fitness model introduced in
Ref.~\cite{gcalda03}. In Sec.~\ref{sec:pract-algor-gener} we present
an algorithm to generate correlated random networks with a given
\textit{a priori} correlation structure. Sec.~\ref{sec:non-equil}
deals with the mapping into this class of models of non-equilibrium
growing networks. Finally, in Sec.~\ref{sec:conclusions} we draw our
conclusions and perspectives.

\section{Measuring correlations in complex networks}
\label{sec:meas-corr-compl}

\subsection{Two vertices correlations}

Let us consider the class of unstructured undirected networks, in
which all vertices with the same degree can be considered to be
statistically equivalent. In this sense, the following results will
not apply to structured networks, in which a distance ordering can be
defined; for instance, when the small-world property is absent
\cite{klemm02,morenostructured}. A network is said to be
\textit{uncorrelated} when the probability that an edge departing from
a vertex of degree $k$ arrives at a vertex of degree $k'$ is
independent of the degree of the initial vertex $k$.  Most natural
networks are not uncorrelated, in the sense that the degrees at the
end points of any given edge are not independent.  This kind of two
vertices degree correlations can be measured in undirected networks by
means of the conditional probability $\condPN{k'}{k}$ that a vertex of
degree $k$ is connected to a vertex of degree $k'$. From the point of
view of correlations, it is useful to consider the restricted subset
of undirected \textit{Markovian} random networks \cite{marian1}, which
are completely defined by the degree distribution $P(k)$ and the
conditional probability $\condPN{k'}{k}$.  The Markovian nature of
this class of networks implies that all higher order correlations can
be expressed as a function of $\condPN{k'}{k}$.

The functions $P(k)$ and $\condPN{k'}{k}$ are assumed to be
normalized, i.e.
\begin{equation}
  \sum_k P(k) = \sum_{k'} \condPN{k'}{k} =1,
  \label{eq:2}
\end{equation}
and they are constrained by a degree detailed balance condition
\cite{marian1} stating the physical conservation of edges among
vertices: The total number of edges pointing from vertices with degree
$k$ to vertices with degree $k'$ must be equal to the number of edges
that point from vertices $k'$ to vertices $k$.  There is an intuitive
way to derive the degree detailed balance condition \cite{marianproc}.
Let us denote by $N_k$ the number of vertices of degree $k$. Since
$\sum_k N_k =N$, where $N$ is the size of the network, we can define the
degree distribution as $P(k) = N_k /N$ \footnote{Being more precise,
  we should write $P(k) \equiv \lim_{N\to \infty} N_k / N$, although, for the
  sake of simplicity, we will obviate the limit here and in the
  following.}. To completely define the network, we need to specify
also how the different degree classes are connected. To this end, let
us define the symmetric matrix $E_{k k'}$, that gives the number of
edges between vertices of degree $k$ and $k'$, for $k \neq k'$, and two
times the number of self-connections for $k=k'$ (the number of
connections between vertices in the same degree class).  This matrix
fulfills the identities
\begin{eqnarray}
  \sum_{k'} E_{k k'} &=& k N_k, \\
  \sum_{k, k'} E_{k k'} &=& \avk N = 2 E,
\end{eqnarray}
where $E$ is the total number of edges in the network. This last
identity allows to define the \textit{joint distribution}
\begin{equation}
  P(k,k')=\frac{E_{k k'}}{\langle k \rangle N},
\end{equation}
where the symmetric function $(2-\delta_{k,k'})P(k,k')$ is the probability
that a randomly chosen edge connects two vertices of degrees $k$ and
$k'$.  The conditional probability, $\condPN{k'}{k}$, defined as the
probability that an edge from a $k$ vertex points to a $k'$ vertex,
can be easily written as
\begin{equation}
  \condPN{k'}{k} = \frac{E_{k'   k}}{k N_k} =
  \frac{\avk P(k,k')}{k P(k)}.
  \label{condP}
\end{equation}
From the symmetry of $P(k,k')$ it follows immediately the degree
detailed balance condition
\begin{equation}
  k \condPN{k'}{k} P(k) = k' \condPN{k}{k'}  P(k') = \avk P(k,k').
  \label{eq:3}
\end{equation}
The joint distribution, $P(k,k')$, conveys all the information needed
to construct a Markovian random network. In fact, it is easy to see
that
\begin{equation}
  P(k) = \frac{\avk}{k} \sum_{k'} P(k,k').
  \label{eq:1}
\end{equation}
This relation, together with Eq.~(\ref{condP}), completely defines the
network properties, i.e. $P(k)$ and $\condPN{k'}{k}$, as a function of
the joint distribution $P(k,k')$. Notice that Eqs.~(\ref{condP}) and
~(\ref{eq:1}) define the degree distribution and the conditional
probability in the whole $k$ range, except for $k=0$. These values can
be recovered by noticing that  $\condPN{k'}{0} \equiv 0$, and, from the
normalization condition Eq.~(\ref{eq:2}),
\begin{equation}
  P(0) = 1 - \sum_{k \geq 1}  \frac{\avk}{k} \sum_{k'} P(k,k').
\end{equation}

The empirical evaluation of $P(k,k')$ (or $\condPN{k'}{k}$) is, in
most real networks, a quite difficult task, since the available data,
restricted to finite sizes, usually yields results extremely noisy and
difficult to interpret. For this reason, it is more useful for
practical purposes to analyze instead the average degree of the
nearest neighbors (ANND) as a function of the vertex degree, defined
by \cite{alexei}
\begin{equation}
  \bar{k}_{nn}(k) = \sum_{k'} k' \condPN{k'}{k}.
  \label{eq:7}
\end{equation}
For uncorrelated networks, in which $\condPN{k'}{k}$ does not depend
on $k$, application of the normalization condition Eq.~(\ref{eq:2})
into Eq.~(\ref{eq:3}) yields $P_0(k' \, \vert \, k) = k' P(k') /
\avk$. In this case, we obtain $ \bar{k}_{nn}^0(k) = \fluck / \avk$,
independent of $k$. Therefore, a function $\bar{k}_{nn}(k)$ with an
explicit dependence on $k$ signals the presence of degree correlations
in the network. Based on the ANND, it is possible to characterize the
correlation properties of the network \cite{assortative}: When
$\bar{k}_{nn}(k)$ is an increasing function of $k$, the network shows
assortative mixing. Examples of assortative behavior can be found in
several social networks \cite{assortative}. On the other hand, when
$\bar{k}_{nn}(k)$ is a decreasing function of $k$, the network shows
disassortative mixing, as found for example is technological systems
such as the Internet \cite{alexei}.

\subsection{Three vertices correlations}

Correlations among three vertices can be measured by means of the
probability $P(k', k'' \, \vert \, k)$ that a vertex of degree $k$ is
simultaneously connected to two vertices with degrees $k'$ and $k''$.
In the particular case of Markovian networks, this function is related
to the two vertices correlation through $P(k', k'' \, \vert \, k) =
\condPN{k'}{k} \condPN{k''}{k}$. For non Markovian networks, however,
the functions $P(k', k'' \, \vert \, k)$ and $\condPN{k'}{k}$ are in
principle not related.

Information about three vertices correlations can be obtained from the
clustering coefficient. The concept of clustering in a graph refers to
the tendency to form cliques (complete subgraphs \cite{bollobas98}) in
the neighborhood of any given vertex. In this sense, clustering
implies that if the vertex $i$ is connected to the vertex $j$, and at
the same time $j$ is connected to $l$, then, with high probability,
$i$ is also connected to $l$. The probability that two vertices with a
common neighbor are also connected to each other is called the
\textit{clustering coefficient} of the common vertex \cite{watts98}.
Numerically, the clustering coefficient, $c_i$, of the vertex $i$ can
be computed as the ratio between the number of edges existing between
the $k_i$ neighbors of $i$, $e_i$, and its maximum possible value,
$k_i(k_i-1)/2$, that is,
\begin{equation}
  c_i = \frac{ 2 e_i}{k_i(k_i-1)}.
\end{equation}
On the other hand, the clustering coefficient of a vertex of degree
$k$, $\bar{c}(k)$ \cite{alexei02}, can be formally computed as the
probability that it is connected to vertices $k'$ and $k''$, and that
those two vertices are, on their turn, joined by and edge, averaged
over all the possible values of the degrees of the neighbor vertices.
Therefore, we can write $\bar{c}(k)$ as a function of the three
vertices correlations as
\begin{equation}
  \bar{c}(k)  = \sum_{k', k''} P(k', k'' \, \vert \, k) p_{k', k''},
  \label{eq:29}
\end{equation}
where the function $p_{k', k''}$ is the probability that the vertices
$k'$ and $k''$ are connected \footnote{Note that the probability
  $p_{k', k''}$ will in general depend on the particular network
  considered, and can also be a function of the degree $k$ of the
  common vertex.}.  The quantity $\bar{c}(k)$ has been recently used
to study the level of hierarchy and modularity in real complex
networks \cite{ravasz02}.

\section{Hidden variable models of  correlated networks}
\label{sec:hidd-vari-models}

Recently, Caldarelli \textit{et al.} \cite{gcalda03} and S{\"o}derberg
\cite{soderberg} (see also Ref.~\cite{goh01}) have proposed different
models of inhomogeneous random graphs that represents a natural
generalization of the classical Erd{\"o}s-R{\'e}nyi random graph model
\cite{erdos60,gilbert59}.  These models consider inhomogeneous graphs
in which each vertex is characterized by a different \textit{type} or
\textit{fitness}. Types can be either discrete or continuous variables
and are assigned to vertices according to a certain probability
distribution. Then, pairs of vertices are independently joined by an
undirected edge with a probability depending on the type of the
respective end points. This construction leads to an ensemble of
undirected random networks, which inherits the simplicity of the
Erd{\"o}s-R{\'e}nyi model while allowing freedom for general forms of the
degree distribution and correlation structure.  Ref.~\cite{soderberg}
was mainly concerned with the component distribution and the onset of
the giant component in this kind of models, and Ref.~\cite{gcalda03}
reported numerical simulations for different model parameters, and
analytical arguments for the form of the degree distribution.

The models defined in Refs.~\cite{gcalda03,soderberg} can be
generalized as a class of models with \textit{hidden variables}. The
hidden variables play the role of tags assigned to the vertices, and
they completely determine the topological properties of the network
through their probability distribution and the probability to connect
pairs of vertices.

We define the class of models with hidden variables as follows: Let us
consider a set of $N$ disconnected vertices and a general hidden
variable $h$, that can be a natural or a real number. An undirected
graph is generated by the following two rules:

\begin{enumerate}
\item Each vertex $i$ is assigned a variable $h_i$, independently
  drawn from the probability distribution $\rho(h)$.
\item For each pair of vertices $i$ and $j$, with respective hidden
  variables $h_i$ and $h_j$, an undirected edge is created with
  probability $r(h_i, h_j)$ (the \textit{connection probability}),
  where $r(h,h') \geq 0$ is a symmetric function of $h$ and $h'$.
\end{enumerate}

Given the independent assignment of hidden variables and edges among
vertices, this procedure generates correlated random networks with
neither loops nor multiple edges, which are Markovian at the hidden
variable level \footnote{But not generally at the actual degree
  level.}  and whose degree distribution and correlation properties
are encoded in the two functions $\rho(h)$ and $r(h,h')$. Here we will
focus in the case in which the distribution $\rho(h)$ is
\textit{independent} of the network size $N$. The case in which
$\rho(h)$ is allowed to depend on $N$ will be considered in
Sec.~\ref{sec:non-equil}.

In this Section we will provide analytic expressions for the
correlation function and clustering coefficient of the networks
generated with this class of models as a function of the distribution
of hidden variables and the probability to connect pairs of vertices.

\subsection{Degree distribution}

The degree distribution, $P(k)$, is defined as the probability that any
given vertex has $k$ edges attached to it. Therefore, in order to
compute it, we need to know the conditional probability $\prop{k}{h}$
(propagator) that a vertex with initial hidden variable $h$ ends up
connected to other $k$ vertices. The degree distribution can then be
written as
\begin{equation}
  P(k) = \sum_h \prop{k}{h} \rho(h),
\label{eq:6}
\end{equation}
where the summation sign must be exchanged by an integral for
continuous $h$. The propagator, which is obviously normalized,
$\sum_k \prop{k}{h} =1$, provides full information about the
dependence of the actual degree $k$ on the hidden variable $h$. In
particular, we can see that the average degree of the vertices
with hidden variable $h$, $\bar{k}(h)$, is given by
\begin{equation}
  \bar{k}(h) = \sum_k k \prop{k}{h},
\end{equation}
and the average degree can be expressed as
\begin{equation}
  \avk = \sum_k k P(k) = \sum_h  \bar{k}(h)  \rho(h).
  \label{eq:27}
\end{equation}
On the other hand, the probability that a vertex of actual degree $k$
has associated a hidden variable $h$, $g^*(h \, \vert \, k)$, can be
computed as the inverse of the propagator by means of Bayes' formula
\cite{gnedenko},
\begin{equation}
  P(k) g^*(h \, \vert \, k) = \rho(h) \prop{k}{h}.
  \label{eq:34}
\end{equation}

In order to get an explicit expression for the propagator, we start by
noticing that it can be written as
\begin{widetext}
  \begin{equation}
  \prop{k}{h} = \sum_{k_1 ,\ldots, k_c} \pprop{1}{k_1}{h_1}
  \pprop{2}{k_2}{h_2} \cdots  \pprop{c}{k_c}{h_c} \delta_{k_1+k_2+\cdots +k_c, k} \label{eq:4},
\end{equation}
\end{widetext}
where $\pprop{i}{k_i}{h_i}$ is the probability that a vertex with
hidden variable $h$ ends up with $k_i$ connections with vertices of
hidden variable $h_i$, $h_c$ being the maximum value of $h$.  Since
the connections between vertices with hidden variables $h$ and $h'$
are independently drawn with probability $r(h,h')$, the probability
$\pprop{i}{k_i}{h_i}$ is simply given by a binomial distribution, i.e.
\begin{equation}
  \pprop{i}{k_i}{h_i} =\binom{N_i}{k_i} r(h, h_i)^{k_i} \left[ 1-
    r(h, h_i)\right]^{N_i - k_i},
\end{equation}
where $N_i = N \rho(h_i)$ is the number of vertices with hidden variable
$h_i$. Let us define now the generating function \cite{wilf94}
\begin{equation}
  \tprop{z}{h} = \sum_k z^k \prop{k}{h}.
  \label{eq:26}
\end{equation}
Since the propagator is given by a convolution, Eq.~(\ref{eq:4}), we
can write its generating function as the product of the generating
functions of the partial propagators $\pprop{i}{k_i}{h_i}$, which on
their turn, being binomial distributions, yield
\begin{equation}
  \tpprop{i}{z}{h_i} = \left[ 1- (1-z)r(h, h_i) \right]^{N_i}.
\end{equation}
Inserting this expression into the definition of $\tprop{z}{h}$ and
taking logarithms on both sides we are led to the equation
\begin{equation}
  \ln \tprop{z}{h} = N \sum_{h'} \rho(h') \ln \left[ 1- (1-z)r(h, h')
  \right].
\label{eq:5}
\end{equation}
For general probabilities $\rho(h)$ and $r(h, h')$,
Eq.~(\ref{eq:5}) must be solved and inverted in order to obtain
the corresponding propagator. The degree distribution is then
obtained applying Eq.~(\ref{eq:6}). Even without solving the
previous equation, however, it is already possible to obtain some
information on the connectivity properties of the network. From
the definition Eq.~(\ref{eq:26}), the first moment of
$\prop{k}{h}$ is given by the first derivative of $\tprop{z}{h}$
evaluated at $z=1$. Therefore we have
\begin{equation}
  \bar{k}(h) = N \sum_{h'} \rho(h') r(h, h'),\label{eq:31}
\end{equation}
and
\begin{equation}
  \avk = N\sum_{h, h'} \rho(h)  r(h, h')  \rho(h'), \label{eq:28}
\end{equation}
where we have used Eq.~(\ref{eq:5}) in computing these expressions.

At this point we must consider the possibility of two different
kinds of networks: Sparse networks, with a well-defined
thermodynamic limit for the average degree $\avk$, and non-sparse
networks, in which the average degree diverges with the network
size. In the case of sparse networks, the number of edges grows
linearly with the system size and, therefore, the joint
distribution is a well defined quantity, independent of $N$.  In
the opposite case, non-sparse networks have a number of edges
growing faster that linearly, which causes the breakdown of the
thermodynamic limit and the emergence of the phenomenon of
condensation of edges (see Sec.~\ref{sec:intr-fitn-model}). In
order to distinguish between sparse and non-sparse networks we
must consider the value of the average degree, given by
Eq.~(\ref{eq:28}).  If the density $\rho(h)$ is independent of the
size of the system, the only possibility to have a sparse network
is that the connection probability scales as $N^{-1}$. This
scaling behavior turns out to have a strong implication in the
form of the propagator. Defining $r(h, h') \equiv C(h,h')/N$ (as
considered in Ref.~\cite{soderberg}), where $C(h,h')$ is a bounded
symmetric function, independent of $N$, we can expand the
right-hand-side of Eq.~(\ref{eq:5}) in the limit $N\to\infty$ to
obtain
\begin{equation}
  \tprop{z}{h} = \exp \left\{ (z-1) \sum_{h'}  \rho(h')  C(h,h') \right\}.
  \label{eq:32}
\end{equation}
The generating function of the propagator is a pure exponential, which
indicates that the propagator itself is a Poisson distribution,
\begin{equation}
  \prop{k}{h} = \frac{e^{-\bar{k}(h)} \bar{k}(h)^k}{k!},
  \label{eq:19}
\end{equation}
where in this case $\bar{k}(h) = \sum_h \rho(h') C(h,h')$.  Eq.~(\ref{eq:19})
is, indeed, a strong result since is states the universality of the
propagator for sparse networks regardless of the form of the
connection probability.

\subsection{Degree correlations}

Degree correlations are completely characterized by means of the
conditional probability $\condPN{k'}{k}$, that gives the
probability that an edge emanating from a vertex of degree $k$ is
connected to a vertex of degree $k'$.  In order to construct the
function $\condPN{k'}{k}$ we consider a vertex of degree $k$, that
with probability $g^*(h \, \vert \, k)$ has associated a hidden
variable $h$. Let us define $\condpN{h'}{h}$ the conditional
probability that a $h$ vertex is connected to a $h'$ vertex. Then,
the conditional probability  $\condPN{k'}{k}$ can be written as
\begin{equation}
\condPN{k'}{k} = \sum_{h, h'} \prop{k'-1}{h'} \condpN{h'}{h} g^*(h
\, \vert \, k),
\end{equation}
where the propagator $\prop{k'-1}{h'}$ gives the probability that the
$h'$ vertex ends up with degree $k'$ (since one connection has already
been used up for the conditional edge with $h$).  Using the form of
$g^*(h \, \vert \, k)$ given by Eq.~(\ref{eq:34}) we have
\begin{equation}
\condPN{k'}{k} =\frac{1}{P(k)} \sum_{h,h'} \prop{k'-1}{h'}
\condpN{h'}{h}  \rho(h)   \prop{k}{h}, \label{eq:8bis}
\end{equation}
valid for $k, k' = 1, 2, \ldots$.  In order to close Eq.~(\ref{eq:8bis}),
we need, finally, to provide an expression for the conditional
probability $\condpN{h'}{h}$. In order to do so, we consider that the
probability of drawing an edge from $h$ to $h'$ is proportional to the
probability of finding an $h'$ vertex, times the probability of
creating the actual edge. Taking into account normalization, we have
that
\begin{equation}
  \condpN{h'}{h} = \frac{\rho(h') r(h, h')}{\sum_{h''} \rho(h'') r(h, h'')}=
  \frac{N \rho(h') r(h, h')}{\bar{k}(h)}.
  \label{eq:14}
\end{equation}
Finally, using Eqs.~(\ref{eq:8bis}) and~(\ref{eq:14}) we can
compute the ANND as
\begin{equation}
  \bar{k}_{nn}(k) = 1+\frac{1}{P(k)} \sum_{h} \prop{k}{h}
  \rho (h) \bar{k}_{nn}(h),
\label{eq:13}
\end{equation}
where we have defined the ANND of a $h$ vertex (see
Eq.~(\ref{eq:7})) as
\begin{equation}
\bar{k}_{nn}(h)\equiv \sum_{h'} \bar{k}(h')
p(h'|h).\label{knnhidden}
\end{equation}

\subsection{Clustering coefficient}

The clustering coefficient is defined as the probability that two
vertices, adjacent to a third vertex, are also connected to each
other. In the space of hidden variables, consider a $h$ vertex,
which is connected to two other vertices $h'$ and $h''$ with
probability $p(h', h'' \, \vert \, h)$. On the other hand, $h'$
and $h''$ are connected with probability $r(h', h'')$. Therefore,
the clustering coefficient of a vertex $h$ in given by $c_h =
\sum_{h', h''}  p(h', h'' \, \vert \, h) r(h', h'')$. Note that
this is the natural counterpart of Eq.~(\ref{eq:29}) in the space
of hidden variables. Now, since the network is Markovian at the
hidden variable level, we have that $p(h', h'' \, \vert \, h) =
\condpN{h'}{h} \condpN{h''}{h}$. Thus we have that
\begin{equation}
  c_h = \sum_{h', h''} \condpN{h'}{h} r(h',  h'') \condpN{h''}{h}.
  \label{eq:21}
\end{equation}
The clustering coefficient of the vertices of degree $k$,
$\bar{c}(k)$, will be given by the probability that a vertex $k$ has
hidden variable $h$, $g^*(h \, \vert \, k)$, times $c_h$, averaged
over all the possible values of $h$. Thus
\begin{equation}
  \bar{c}(k) =\frac{1}{P(k)} \sum_h \rho(h) \prop{k}{h} c_h, \quad k = 2,
3, \ldots,
    \label{eq:8}
\end{equation}
where we have used the form of $g^*(h \, \vert \, k)$ given by
Eq.~(\ref{eq:34}).

The results derived in this Section represent the general solution of
the class of networks with hidden variables. In the rest of the paper
we will show how this formalism is able to deal with a wide variety of
models, from sparse to non-sparse networks, and from equilibrium to
non-equilibrium ones.

\section{The intrinsic fitness model}
\label{sec:intr-fitn-model}

As an example of the general class of models with hidden variables,
Caldarelli \textit{et al.} \cite{gcalda03} considered the model
defined by the probability distributions
\begin{eqnarray}
  \rho(h) &=& e^{-h} \qquad \mathrm{for} \; h\in [0,\infty[, \label{eq:9}\\
  r(h, h') &=& \theta(h + h' -\zeta), \label{eq:10}
\end{eqnarray}
where $\theta(x)$ is the Heaviside step function and $\zeta$ is a constant.
In this model, hereafter referred to as the intrinsic fitness (IF)
model, vertices have assigned an exponentially distributed hidden
variable (fitness), and are joined by an edge whenever the sum of the
fitness of the end points is larger than a given threshold $\zeta$. By
means of numerical simulations and analytical arguments, Caldarelli
\textit{et al.} \cite{gcalda03} showed that the degree distribution in
this model is power-law distributed. This observation led to the very
interesting conclusion that it is possible to generate scale-free
networks without growth not preferential attachment \cite{barab99}.

\subsection{Analytic solution}

Using the general formalism developed in the previous Section, we can
provide analytic expressions for the main properties of the IF model.
In order to do so, let us compute in the first place the propagator
$\prop{k}{h}$.  Inserting Eqs.~(\ref{eq:9}) and~(\ref{eq:10}) into
Eq.~(\ref{eq:5}), we have, substituting the summation by an integral,
\begin{widetext}
\begin{equation}
   \ln \tprop{z}{h} = N \int_0^\infty d h' e^{-h'} \ln \left[ 1- (1-z) \theta(h
     + h' -\zeta)  \right] = N \ln z \begin{cases}
                                 e^{h -\zeta} & \mathrm{if} \; 0 \leq h \leq \zeta \\
                                 1 & \mathrm{if} \; h > \zeta
                                 \end{cases},
\end{equation}
from where we obtain $\tprop{z}{h} = z^{N e^{h-\zeta}}$ for $ 0 \leq h \leq
\zeta$, and $\tprop{z}{h} = z^N$ for $h > \zeta$. In order to invert this
generating function, we approximate $k$ by a continuous variable.  In
this case, the propagator takes the simple form
\begin{equation}
  \prop{k}{h} = \delta(k - N  e^{h-\zeta}) \theta_h(0, \zeta) + \delta(k-N) \theta(h-\zeta),
\label{eq:11}
\end{equation}
where $\delta(x)$ is the Dirac delta function and we have
introduced the window function
\begin{equation}
  \theta_x(a, b) = \begin{cases}
             1 & \quad \mathrm{for} \;  a \leq x \leq b \\
             0 & \quad \mathrm{otherwise}
             \end{cases}.
\end{equation}
This approximation is expected to perform poorly for small values of
$k$, as we will see when comparing the analytical results with
computer simulations of the IF model.

Inserting the propagator Eq.~(\ref{eq:11}) into the general expression
Eq.~(\ref{eq:6}), and performing the integrals corresponding to the
Dirac delta functions, we obtain the degree distribution
\begin{equation}
  P(k) = N e^{-\zeta} \frac{1}{k^2} \theta_k(N e^{-\zeta}, N) +  e^{-\zeta}
  \delta(k-N).
  \label{eq:12}
\end{equation}
That is, the networks generated by the IF model exhibit a scale-free
degree distribution, with degree exponent $\gamma=2$, for degrees in the
range $Ne^{-\zeta} \leq k \leq N$, plus an accumulation point at $k=N$, given
by the delta function, with weight $e^{-\zeta}$.  This accumulation point
signals the presence of a \textit{condensation} of edges in the
fraction $e^{-\zeta}$ of the vertices of the network with $h>\zeta$, that
establish connections to all the other vertices \cite{bianconi01}.
This condensation, reminiscent to that observed in models with
non-linear preferential attachment \cite{krap00}, is the result of the
non-sparse nature of the network, that, from Eq.~(\ref{eq:28}), has
average degree $\avk = N e^{-\zeta} (\zeta+1)$.

In order to characterize the correlations of the model, we compute
the ANND, given by Eq.~(\ref{eq:13}). In the continuous $k$
approximation, the function $\bar{k}(h)$ takes the form
\begin{equation}
  \bar{k}(h)=N e^{h-\zeta}  \theta_h(0, \zeta) + N \theta(h-\zeta).
  \label{eq:30}
\end{equation}
Inserting this expression into the formula for the ANND, we obtain
\begin{equation}
\bar{k}_{nn}(k)=1+\frac{N e^{-2\zeta}}{P(k)}
\left[(1+\zeta)\delta(k-N)+\frac{N^2}{k^3}
\left\{ 1+\zeta+\ln{\left(\frac{k}{N}\right)}\right\}\theta_k(N e^{-\zeta}, N) \right]
\end{equation}
The regular part of this expression (discarding the delta function
singularities, signaling again the effect of the condensation of
edges in the correlation function) takes the form
\begin{equation}
  \bar{k}_{nn}^r(k) = 1 + \frac{N^2 e^{-\zeta}}{k} \left[ 1+\zeta +
    \ln\left(\frac{k}{N}\right) \right] \theta_k(N e^{-\zeta}, N).
\end{equation}
That is, the regular part of the ANND is proportional to $k^{-1}$,
times a logarithmic correction term. We are therefore in the presence
of disassortative mixing. Note that, in the limit $N\to\infty$, we have that
$ \bar{k}_{nn}^r(k) \to \infty$, in agreement with the theoretical
prediction made in Ref.~\cite{marian3}.

Finally, to estimate the clustering coefficient, we have to compute
first the conditional probability at the level of hidden variables,
given by Eq.~(\ref{eq:14}). Using Eqs.~(\ref{eq:9}) and~(\ref{eq:10}),
we obtain
\begin{equation}
  \condpN{h'}{h} = e^{-h'} \theta(h'+h-\zeta) \left[ e^{\zeta-h} \theta_h(0,\zeta) +
    \theta(h-\zeta) \right].
\end{equation}
From this expression we can obtain the clustering coefficient at
the level of the hidden variables
\begin{equation}
  c_h =  \theta_h(0, \zeta/2) + e^{\zeta-2h} (2h-\zeta+1)  \theta_h(\zeta/2, \zeta) +e^{-\zeta}
  (\zeta+1) \theta(h-\zeta),
\end{equation}
and the clustering coefficient as a function of the degree $k$,
\begin{eqnarray}
  \bar{c}(k) &=& \frac{N e^{-\zeta} }{k^2 P(k)} \theta_k(Ne^{-\zeta}, Ne^{-\zeta/2})
  + \frac{N^3 e^{-2\zeta}}{k^4 P(k)} \left[ 2
    \ln\left(\frac{k}{N} \right) +\zeta+1 \right]  \theta_k(Ne^{-\zeta/2}, N)
  \nonumber\\
  && +\frac{1}{P(k)} e^{-2\zeta} (\zeta+1) \delta(k-N).
\end{eqnarray}
The regular part of this formula is finally
\begin{equation}
  \bar{c}^r(k) =  \theta_k(Ne^{-\zeta}, Ne^{-\zeta/2}) + \frac{N^2 e^{-\zeta}}{k^2}
  \left[ 2 \ln \left(\frac{k}{N}\right) +\zeta+1\right]  \theta_k(Ne^{-\zeta/2},
  N).
\end{equation}
That is, for $k\leq Ne^{-\zeta/2}$ the clustering coefficient is constant
and equal to its maximum possible value $1$. The presence of this flat
region in the clustering coefficient is easy to understand. The degree
range $k\leq Ne^{-\zeta/2}$ corresponds, from Eq.~(\ref{eq:30}), to
vertices with fitness $h< \zeta /2$.  These vertices can only establish
connections with vertices with $h' > \zeta/2$, which are on their turn
fully interconnected among them. From here, it follows a maximum
clustering coefficient equal to $1$ for all vertices with $h< \zeta /2$.
On the other hand, for $ Ne^{-\zeta/2} \leq k \leq N$, the clustering
coefficient decreases as $k^{-2}$, modulated again by a logarithmic
correction term.

\subsection{Numerical simulations}

In order to check the validity of the proposed analytical expressions,
we have performed numerical simulations of the IF model. To simplify the
comparison, we have considered the particular case $\zeta=\ln N$, in which
the relevant expressions take the form:
\begin{eqnarray}
  P(k) &=& \frac{1}{k^2} \theta_k(1, N) + \frac{1}{N} \delta(k-N)\label{eq:15}
  \\
  \bar{k}_{nn}^r(k) &=& \left[ 1 + N \frac{1+\ln k}{k}\right]  \theta_k(1,
  N)\label{eq:16}\\
   \bar{c}^r(k) &=& \theta_k(1, N^{1/2}) + \frac{2 N}{k^2} \left[\ln
     \left( \frac{k}{N^{1/2}} \right)+ \frac{1}{2} \right]
   \theta_k(N^{1/2}, N)\label{eq:17}
\end{eqnarray}
Simulations were performed for networks of size $N=10^4$
(corresponding to $\zeta=\ln 10^4 \approx 9.2103$), averaging all statistical
distributions over $10^3$ to $10^5$ network realizations.

In Fig.~\ref{fig:pk} we depict the results corresponding to the
degree distribution. As we can see, the theoretical prediction (solid
line) overestimates the value of the actual degree distribution for
small $k$. This is a natural effect of the continuous $k$
approximation, that can be readily understood from the form of
Eq.~(\ref{eq:15}): The form $P(k) \sim k^{-2}$ cannot be correct in a
discrete approximation, since it does not fulfill the normalization
condition. The condensation of edges at $k=N$ is clearly visible in
the presence of an isolated peak, of height approximately equal to
$N^{-1}$. Figs.~\ref{fig:Knn} and~\ref{fig:Ck}, on the other hand,
represent the ANND and the clustering coefficient as a function of the
degree $k$, respectively. As we can see, the fit between the computer
simulations and the analytical expressions is quite good. \vspace*{0.5cm}
\end{widetext}

\begin{figure}[t]
  \epsfig{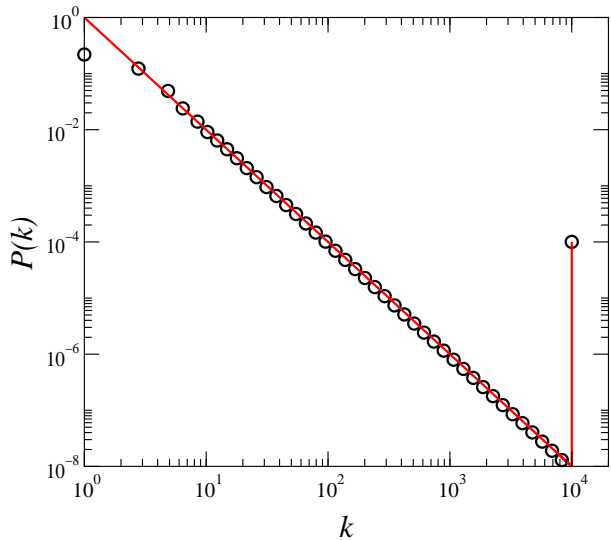}
  \caption{Comparison between the theoretical prediction
    Eq.~(\protect\ref{eq:15}) for the degree distribution (solid line)
    and computer simulations (hollow circles) of the IF model. The
    isolated point at $k=N$ corresponds to the analytical Dirac delta
    function, with strength $e^{-\zeta} = N^{-1}$.}
  \label{fig:pk}
\end{figure}

\begin{figure}[t]
  \epsfig{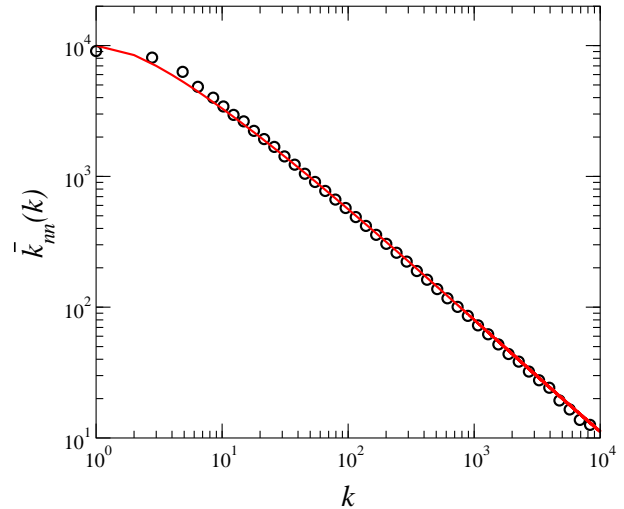}
  \caption{Comparison between the theoretical prediction
    Eq.~(\protect\ref{eq:16}) for the ANND (solid line) and computer
    simulations (hollow circles) of the IF model.}
  \label{fig:Knn}
\end{figure}

\begin{figure}[t]
  \epsfig{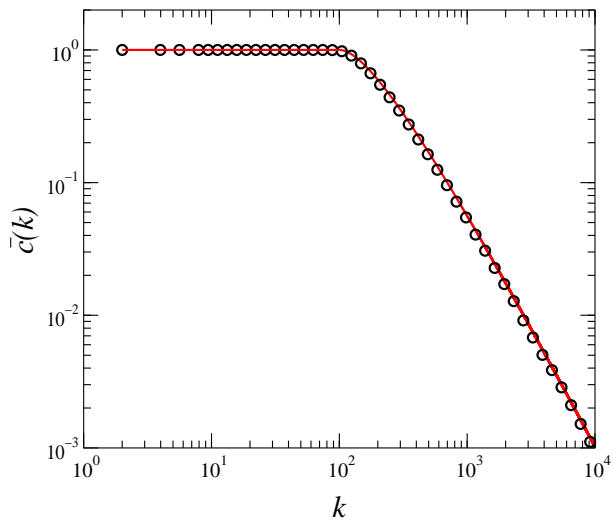}
  \caption{Comparison between the theoretical prediction
    Eq.~(\protect\ref{eq:17}) for the clustering coefficient as a
    function of the degree $k$ (solid line) and computer simulations
    (hollow circles) of the IF mode.}
  \label{fig:Ck}
\end{figure}

\section{A practical algorithm to generate correlated random networks}
\label{sec:pract-algor-gener}

The hidden variable class of models represents a natural extension of
the Erd{\"o}s-R{\'e}nyi random graph model, that allows to generate a broad
class of correlated networks from which it is possible to compute the
most relevant topological properties.  From a practical point of view,
however, it is still missing an important point. Indeed, there are
many situations in which it is desirable to generate a network with a
particular correlation structure given by a certain joint distribution
$P(k, k')$. For a hidden variable model it is possible to compute this
quantity as a function of the initial probabilities $\rho(h)$ and
$r(h,h')$. However, this relation is non-trivial, and it is generally
not possible to invert it.  Therefore, in order to implement an
algorithm capable to generate networks with any \textit{a priori}
correlation structure one must carefully choose the distribution of
hidden variables and the connection probability.

One possible way to proceed is to define hidden variables $h$ that
have themselves the structure of the degrees of a real network
(\textit{hidden degrees}), with correlations given by a joint
distribution $\tilde{P}(h, h')$.  Those hidden degrees will then
be natural numbers, that are assigned to the vertices according to
the probability distribution (see Eq.~(\ref{eq:1}))
\begin{equation}
  \rho(h) = \frac{\avh}{h} \sum_{h'} \tilde{P}(h, h'),
\end{equation}
with $\avh = \sum_h h \rho(h)$.  In order to define the connection
probability, we consider that, if the hidden degrees were the
actual degrees characterizing the network, then the total number
of edges between vertices $h$ and $h'$ would be $E_{h  h'} = \avh
\tilde{P}(h, h') N$. Since the total number of $h$ vertices is
$N_h = N \rho(h)$, it is therefore natural to define
\begin{equation}
  r(h, h') = \frac{\avh}{N} \frac{\tilde{P}(h, h')}{\rho(h) \rho(h')}.
  \label{eq:18}
\end{equation}
On the other hand, the conditional probability that a vertex $h$
is connected to a vertex $h'$ is given by (see Eq.~(\ref{condP}))
\begin{equation}
  \condpN{h'}{h} = \frac{\avh \tilde{P}(h, h')}{h \rho(h)}.
\end{equation}
The quantities $\rho(h)$ and $\condpN{h'}{h}$ will be, in this case,
related through the hidden degree detailed balance condition,
\begin{equation}
  h \condpN{h'}{h} \rho(h) =  h' \condpN{h}{h'} \rho(h') = \avh
  \tilde{P}(h, h'),
\end{equation}
as can be checked by inserting into the definition of
$\condpN{h'}{h}$, Eq.~(\ref{eq:14}), the expression Eq.~(\ref{eq:18}).
It is interesting to note that, if two-point correlations are absent
at the level of the hidden variables, then we have that
$\tilde{P}_0(h, h') = h h' \rho(h) \rho(h') / \avh^2$. Thus, the
connection probability reads
\begin{equation}
  r_0(h, h')  = \frac{h h'}{N \avh},
\end{equation}
recovering the model recently introduced by Chung and Lu
\cite{chunglumode} (see also Ref.~\cite{newmanonchunglu}).

Assuming that the connection probability is bounded and decreases
for large network sizes as $N^{-1}$, we can compute the propagator
$\prop{k}{h}$ applying Eq.~(\ref{eq:19}) with $\bar{k}(h) = N
\sum_{h'} \rho(h') r(h, h') = h$, where we have used
Eq.~(\ref{eq:18}). Therefore, the propagator is a simple Poisson
distribution, with average value $h$, i.e.
\begin{equation}
  \prop{k}{h} = \frac{e^{-h} h^k}{k!}.
  \label{eq:20}
\end{equation}
Note that the validity of this result levers on a quite strong
assumption for the boundedness of the connection probability $r(h,
h')$. The nature of this condition is more clearly seen in the
case of uncorrelated networks, with $r_0(h, h') = h h' /N \avh$.
If $r_0(h, h')$ has to decrease as $N^{-1}$, then the maximum
value of the hidden degree must be smaller than $h_c(N) = (N
\avh)^{1/2}$ \cite{chunglumode}, a condition that imposes
restrictions on the maximum degree available for any vertex.

From the propagator Eq.~(\ref{eq:20}), the degree distribution as
a function of $\rho(h)$ follows immediately from Eq.~(\ref{eq:6}):
\begin{equation}
  P(k) = \sum_h \frac{e^{-h} h^k}{k!}  \rho(h).
  \label{eq:23}
\end{equation}
This relation between distributions implies a relation between the
respective moments. Indeed, it is straightforward to prove that
\begin{equation}
  \langle h^n \rangle = \langle k(k-1) \cdots (k-n+1) \rangle,
\end{equation}
and, in particular, the first two moments read
\begin{equation}
  \avh = \avk, \qquad \qquad \fluch = \fluck - \avk.
\end{equation}
It is also instructive to see how we can recover the classical
Erd{\"o}s-R{\'e}nyi random graph model from this formalism. The
Erd{\"o}s-R{\'e}nyi model corresponds to joining pairs of vertices
with a constant probability $p$. Such connection probability
results from imposing uncorrelated hidden degrees with
distribution $\rho_{\mathrm{ER}}(h) = \delta_{\avk ,
  h}$, which yields a Poisson degree distribution with average degree
$\avk$.

In order to compute the ANND function from Eq.~(\ref{eq:13}), we
observe that in this case $\bar{k}(h)=h$. From here we obtain
\begin{equation}
  \bar{k}_{nn}(k) = 1 + \frac{1}{P(k)} \sum_h  \frac{e^{-h} h^k}{k!}  \rho(h)
  \bar{h}_{nn}(h),
  \label{eq:24}
\end{equation}
where, in this case, the average hidden degree of the nearest
neighbors as a function of $h$ (see Eq.~(\ref{knnhidden})) is
\begin{equation}
  \bar{h}_{nn}(h) = \sum_{h'} h' \condpN{h'}{h} = \frac{\avh}{h \rho(h)}
  \sum_{h'} h' \tilde{P}(h, h').
\end{equation}
For uncorrelated networks at the hidden level, the ANND yields
\begin{equation}
  \bar{k}_{nn}^0(k) = 1 + \frac{\fluch}{\avh} = \frac{\fluck}{\avk},
\end{equation}
recovering the well-known result \cite{alexei03}. Finally, the
clustering coefficient takes, from Eq.~(\ref{eq:8}), the form
\begin{equation}
  \bar{c}(k) =\frac{1}{P(k)} \sum_h \frac{e^{-h} h^k}{k!} \rho(h) c_h,
  \label{eq:25}
\end{equation}
where the clustering coefficient in terms of the hidden degrees is
given by
\begin{widetext}
\begin{equation}
  c_h =  \sum_{h', h''} \condpN{h'}{h} r(h',  h'') \condpN{h''}{h} =
  \frac{\avh^3}{h^2 \rho(h)^2 N} \sum_{h', h''}
  \frac{\tilde{P}(h, h') \tilde{P}(h', h'') \tilde{P}(h'', h)}{\rho(h')
    \rho(h'')}.
\end{equation}
\end{widetext}
When correlations are missing in the hidden degree distribution, we
obtain
\begin{equation}
   \bar{c}^0(k) = \frac{\fluch^2}{N \avh^3} = \frac{(\fluck
     -\avk)^2}{N \avk^3},
\end{equation}
recovering the result previously derived by Newman \cite{newmanrev}.

The key point to notice in the above expressions is that the
Poisson propagator Eq.~(\ref{eq:20}) is a sharply peaked function
at $k=h$, that in the large $k$ limit is analogous to a delta
function $\delta_{h,
  k}$. Therefore, in the limit $k\to\infty$, we expect to observe the
behavior
\begin{eqnarray}
  P(k) &\sim & \rho(k), \\
  \bar{k}_{nn}(k) &\sim& 1 + \bar{h}_{nn}(k), \\
  \bar{c}(k) &\sim & c_k.
\end{eqnarray}
That is, the main topological properties referred to the actual degree
$k$ tend to their analogs computed for the hidden degree $h$, with the
sole exception of a constant of order unit added to the ANND function.
We can take advantage of this observation to propose the following
algorithm to generate a correlated random network with theoretical
degree distribution $P_t(k)$ and joint distribution $P_t(k,k')$:
\begin{enumerate}
\item Assign to each vertex $i$ an integer random variable
  $\tilde{k}_i$, $i=1 \cdots N$, drawn from the probability distribution
  $P_t(k)$.
\item For each pair of vertices $i$ and $j$, draw an undirected edge
  with probability $r(\tilde{k}_i, \tilde{k}_j) = \avk P_t(\tilde{k}_i,
  \tilde{k}_j) / N P_t(\tilde{k}_i) P_t(\tilde{k}_j)$.
\end{enumerate}
The outcome of this process will be a random network whose actual
degree structure, in the large $k$ limit, will be distributed
according to the probability $P_t(k)$, with correlations given by
$P_t(k,k')$.

\begin{figure}[t]
  \epsfig{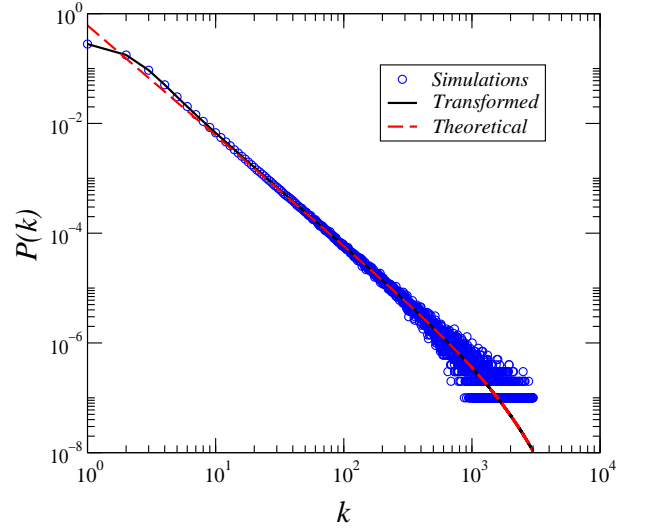}
  \caption{Degree distribution obtained from numerical simulations of
    the proposed algorithm, applied to the joint distribution
    Eq.~(\protect\ref{eq:22}), compared with the theoretical and
    transformed values.}
  \label{fig:pkex}
\end{figure}

\begin{figure}[t]
  \epsfig{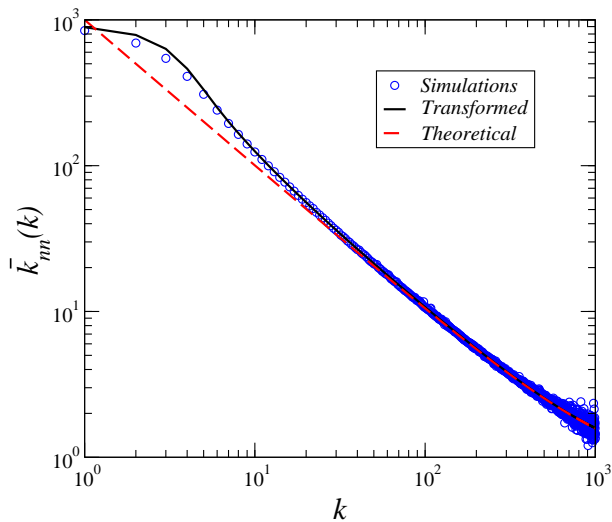}
  \caption{ANND function obtained from numerical simulations of
    the proposed algorithm, applied to the joint distribution
    Eq.~(\protect\ref{eq:22}), compared with the theoretical and
    transformed values.}
  \label{fig:knnex}
\end{figure}

\begin{figure}[t]
  \epsfig{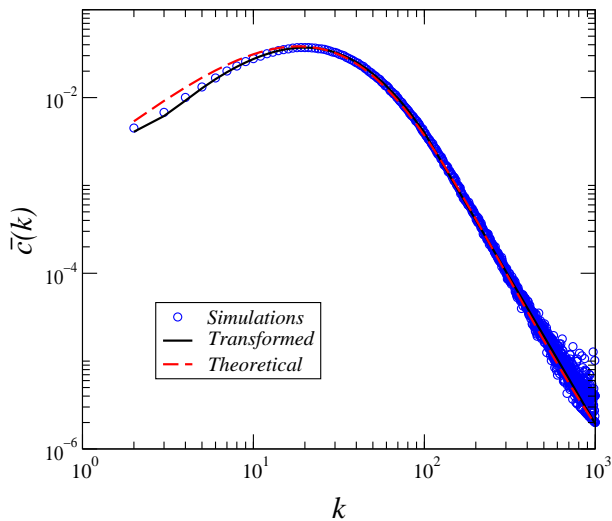}
  \caption{Clustering coefficient obtained from numerical simulations of
    the proposed algorithm, applied to the joint distribution
    Eq.~(\protect\ref{eq:22}), compared with the theoretical and
    transformed values.}
  \label{fig:ckex}
\end{figure}

In order to check the accuracy of the previous algorithm, we have
tested it with the joint probability distribution
\begin{equation}
  P_t(k, k') = A q^{k k'}, \qquad k, k'  =1, 2, \ldots ,
\label{eq:22}
\end{equation}
where $A$ is a normalization constant and $q<1$ is a constant
parameter. With this choice, the degree distribution takes the form
\begin{equation}
  P_t(k) = \frac{\avk A q^k}{k(1-q^k)},
\end{equation}
which, for $q\to1$, approaches a power-law distribution with exponent
$\gamma=2$. In Figs.~\ref{fig:pkex}, \ref{fig:knnex}, and~\ref{fig:ckex}
we present the results for the degree distribution, the ANND function,
and the clustering coefficient, respectively, from computer
simulations of the proposed algorithm, using the joint probability
distribution given by Eq.~(\ref{eq:22}). The plots have been obtained
for networks of size $N=10^4$ and a parameter $q=0.999$, averaging
over $10^3$ realizations. In the same graphs we also represent the
theoretical values corresponding to a network with a correlation
structure given by $P_t(k, k')$, plus the transformed functions given
by Eqs.~(\ref{eq:23}), (\ref{eq:24}), and~(\ref{eq:25}), respectively,
that correspond to the actual topological properties of the network.
As discussed previously, and to ease the comparison of the plots, a
factor $1$ has been subtracted to the ANND function obtained from
computer simulations and the transformation Eq.~(\ref{eq:24}). We can
see that for all three quantities, the matching between the computer
simulations and the theoretical results is very good for values of $k$
larger than $10$.  Being the discrepancy limited to such small degree
values, we conclude that the proposed algorithm reproduces the desired
correlation structure with an accuracy that is more that satisfactory
for any purpose dealing with the large scale properties of the
network.

\section{Non-equilibrium correlated random networks}
\label{sec:non-equil}

The class of models we have introduced so far are static models, in
which, starting from a fixed number $N$ of vertices, edges are
assigned with a given probability. As we have seen, this construction
is extremely useful because it gives us control over the final network
structure and, at the same time, the possibility to calculate
important structural properties. Many real networks, however, are far
from being static. Instead, many of them are the result of an evolving
process \cite{mendesbook}, in which vertices are added to the network
following some growing process (lineal, exponential, etc.),
establishing connections to other existing vertices with a given
attachment rule (preferential attachment \cite{barab99}, deactivation
of vertices \cite{klemm02}, etc).  From this growing mechanism the
network reaches a non-equilibrium steady state where the statistical
properties are time independent.

In the following we will see how it is possible to map non-equilibrium
growing networks into a particular kind of models with hidden
variables.  The key point is to realize that, after the growth of the
network, all vertices that joined the network at the same time are
statistical equivalent and, thus, the hidden variable of a vertex must
correspond to its injection time, $t$ \cite{soderberg}. We consider a
time window $t \in [t_0, T]$, with $T\gg 0$, in which the initial time
can be taken $t_0 =1$ without lack of generality. If $\lambda$ is the rate
of creation of new vertices per unit time \footnote{We consider here a
  linear growth of new vertices.} the network size is given by $N=\lambda
(T-1)$.  In this case, the density of hidden variables, $\rho (t)$, is a
uniform distribution defined in the range $[1,T]$, that is, $\rho (t)=1
/(T-1)$, reflecting the linear growth of the network. The main
difference between this class of non-equilibrium networks and the
classes discussed in the previous Sections lies in the fact that the
distribution $\rho(t)$ has now an explicit dependence on the system size
$N$ (or the final time $T$).

The next step is to define the connection probability of two
vertices that joined the network at times $t$ and $t'$, $r(t,t')$.
The choice of this function is equivalent to the connection
probability at the time that the vertices were added and its
specific form will determine the final properties of the network.
For instance, an homogeneous form for $r(t,t')$, that is,
$r(t,t')=f(|t-t'|)$, will produce, in the large $T$ limit,
networks in which all the vertices will have the same statistical
properties, independent of the injection time.  In the opposite
case of inhomogeneous networks, vertices introduced at early times
will have different topological properties than those added later,
giving rise to \textit{aging} (that is, an explicit dependence on
the injection time $t$ of all the vertex properties evaluated at
time $T > t$).

In order to provide a particular example, we focus on the class of
growing scale-free networks, whose most characteristic element is
the Barab{\'a}si-Albert model \cite{barab99}. From general scaling
arguments \cite{dorogorev,mendesbook}, it is possible to see that
growing scale-free networks are described by a power-law degree
distribution $P(k) \sim k^{-\gamma}$, while the average degree at
time $t$ of a vertex introduced at time $t'$ is given by
\begin{equation}
  k_{t'}(t) \sim \left(\frac{t}{t'}\right)^{\beta} , \qquad 0<\beta <1,
  \qquad t >  t' \gg 1,
  \label{degreeBA}
\end{equation}
where the exponents $\beta$ and $\gamma$ fulfill the scaling relation
\begin{equation}
  \gamma = 1 + \frac{1}{\beta}.
\label{eq:33}
\end{equation}
In the Barab{\'a}si-Albert model, corresponding to $\beta=1/2$ and $\gamma=3$,
new edges are joined to old vertices following a preferential
attachment prescription, that is, with probability proportional to the
degree of the existing vertices. We can generalize this prescription
and consider a preferential attachment as a function of the time $t$,
mapping the degree to the time by means of Eq.~(\ref{degreeBA}). In
this case, however, when a new vertex is added at time $t$, it can
establish connections with nodes introduced between $1$ and $t$ and,
consequently, the connection probability is to be rescaled by a factor
$\int^t k_{t'}(t) dt'$, that is proportional to $t$.  Therefore, the
probability that a new vertex, created at time $t$, will be joined to
a vertex injected at time $t' <t$, is proportional to $(t / t')^{\beta}
/t$.

Following this reasoning we propose a connection probability as a
function of the times $t$ and $t'$ given by
\begin{equation}
r(t,t')=\alpha \left[ \frac{1}{t} \left(
\frac{t}{t'}\right)^{\beta} \Theta(t-t') + \frac{1}{t'} \left(
\frac{t'}{t}\right)^{\beta} \Theta(t'-t) \right],
\label{eq:35}
\end{equation}
where $\alpha$ is a parameter that controls the final average degree of
the network. The connection probability has been symmetrized for $t$
and $t'$ to comply with the general conditions of this function. Its
symmetric property, however, does not imply that the average
properties of the vertices are independent of $t$. For example,
computing the average degree of a $t$ vertex (introduced at time $t$),
evaluated at the final time $T$, using Eq.~(\ref{eq:31}), we obtain
\footnote{The average degree of a vertex injected at time $t$,
  $\bar{k}(t)$, should not be confused with $k_{t}(t')$, that measures
  the average degree of the same vertex at an intermediate time
  $t'>t$.}
\begin{eqnarray}
  \bar{k}(t)&=&\lambda \int_1^T r(t,t')dt' \nonumber\\
  &=&\frac{\alpha
    \lambda}{1-\beta} \left( 1-\frac{1}{t^{1-\beta}}\right)
  +\frac{\alpha \lambda}{\beta} \left( \left( \frac{T}{t}
    \right)^{\beta}-1 \right). \label{condhiddendegree}
\end{eqnarray}
For large $T$, we have $ \bar{k}(t) \sim (T /t)^\beta$, recovering the
behavior obtained in growing network models, Eq.~(\ref{degreeBA}), if
we consider the time $T$ as the observation time.  On the other hand,
$\bar{k}(t)$ is a decreasing function of $t$ between the limits $t=1$,
yielding $\bar{k}_{max} \sim \alpha \lambda T^{\beta} /\beta$, and $t=T$, where it
converges to the constant $\bar{k}_{min} \sim \alpha \lambda /(1-\beta)$.  This
functional form implies that the oldest vertices (with smaller $t$)
have a larger average degree, which is the  signature of aging in the
network \cite{barab99}.

From Eq. (\ref{eq:28}), the average degree of the network can be
computed as
\begin{equation}
  \langle k \rangle=\frac{1}{T-1} \int_1^T
  \bar{k}(t)dt=\frac{2\alpha \lambda}{1-\beta} \left( 1 -\frac{1}{\beta} \frac{T^\beta
      -1}{T-1} \right),
\end{equation}
that, in the limit of large $T$ and $\beta<1$, tends to $\avk = 2 \alpha \lambda
/(1-\beta)$, from where we identify the normalization parameter $\alpha$ as a
function of the average degree.  For $\beta=1$, on the other hand, the
average degree diverges as $\avk \sim 2 \alpha \lambda \ln T$.  Note that this
choice of the connection probability, independent of the network size,
yields for $\beta<1$ a sparse network (with finite average degree in the
thermodynamic limit), in opposition with the case discussed in
Sec.~\ref{sec:hidd-vari-models}. This fact is due to the explicit
dependence on $N$ of the distribution of times $\rho(t)$, and signals
the crucial difference of non-equilibrium networks with hidden
variables.

In order to obtain the form of the degree distribution, we observe
that, even though the network is sparse, since $r(t, t')$ is not
proportional to $N^{-1}$ we cannot rigorously apply
Eq.~(\ref{eq:19}). However, we note that the maximum value of $r(t,
t')$ takes place at $t=t'$, namely
\begin{equation}
  r(t, t) = \frac{\alpha}{t} = \frac{\avk (1-\beta)}{2 \lambda t}.
\end{equation}
For not very large values of $\avk$ and large $t$, the connection
probability is  bounded by an small value, and therefore we can
still approximate its propagator by means of
Eq.~(\ref{eq:19}). Working for simplicity with the generating function
of the degree distribution, $\hat{P}(z) = \sum_k z^k P(k)$, we therefore
write, from Eq.~(\ref{eq:32}),
\begin{equation}
  \hat{P}(z) = \frac{1}{T-1} \int_1^T e^{(z-1)\bar{k}(t)} dt.
\end{equation}
Performing the change of variables $\tau \equiv t / T$, and considering the
limit  $T\to\infty$, we obtain
\begin{widetext}
\begin{eqnarray}
  \hat{P}(z)&=&\exp\left\{\frac{\alpha \lambda (2\beta-1)(z-1)}{ \beta
      (1-\beta)}\right\} \int_0^1 e^{- \alpha \lambda (1-z)
    \tau^{-\beta}/\beta} d\tau \nonumber \\[0.5cm]
  &=&  \left[
    \frac{\alpha \lambda}{\beta} (1-z) \right]^{1/ \beta} \frac{\Gamma(-1/ \beta, \alpha \lambda
    (1-z) / \beta)}{\beta} \exp\left\{\frac{\alpha \lambda (2\beta-1)(z-1)}{ \beta (1-\beta)}\right\},
\end{eqnarray}
\end{widetext}
where $\Gamma(x, y)$ is the incomplete Gamma function. Expanding
$\hat{P}(z)$ around $z=1$, we obtain the leading terms
\begin{equation}
  \hat{P}(z) \simeq 1 - \avk  (1-z) +
    \left(\frac{\alpha \lambda}{\beta} \right)^{1/ \beta} \frac{\Gamma(-1/ \beta)}{\beta}
    (1-z)^{1/ \beta}.
\end{equation}
Applying Tauberian theorems \cite{handelsman}, it follows from the
singularity $ \hat{P}(z) \sim (1-z)^{1/ \beta}$ in the vicinity of $z=1$,
the large $k$ behavior of the degree distribution, namely $P(k) \sim
k^{-1 - 1/ \beta}$.  That is, the model recovers a scale-free network,
$P(k) \sim k^{-\gamma}$, with a degree exponent $\gamma = 1 +1 / \beta$, and a
cut-off given by the maximum degree $\bar{k}_{max} \sim T^{\beta} \sim
N^{1/(\gamma-1)}$, in agreement with the results obtained for growing
network models \cite{mendesbook}.

\begin{figure}[t]
  \epsfig{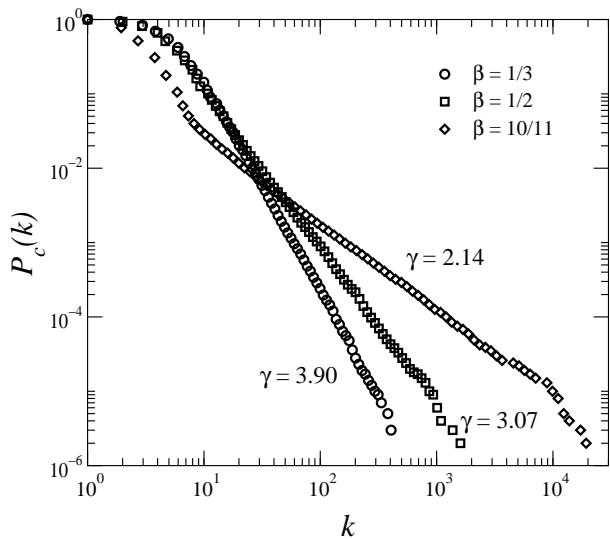}
  \caption{Cumulative distribution for the mapping of the growing model for
  different values of the parameter $\beta$ ($\beta=1/3, 1/2$ and $10/11$).
  The average degree is in all cases set to $\langle k \rangle =6$. The size of the network is
  $N=10^6$.}
  \label{fig:simsnoneq}
\end{figure}

In order to check this result, we have numerically generated networks
of size $N=10^{6}$, with the connection probability Eq.~(\ref{eq:35}).
For a rate of addition of new vertices $\lambda=1$, the numerical prefactor
in $r(t, t')$ is $\alpha = \avk (1-\beta)/2$; we impose $\avk=6$.
Fig.~\ref{fig:simsnoneq} shows the numerical cumulated degree
distributions obtained for values $\beta=1/3$, $1/2$, and $10/11$. The
plots show a clear power-law behavior in all three cases, with a
degree exponent, estimated from a linear fit in the scaling region,
given by $\gamma=3.90$, $3.07$, and $2.14$, respectively. This values
compare very well with the theoretical prediction $\gamma =1 +1/ \beta$, that
provides the expected exponents $4$, $3$, and $2.1$.

The ANND function, $\bar{k}_{nn}(k)$, can also be analyzed using our
formalism. By means of Eq.~(\ref{eq:13}), assuming a Poisson form for
the propagator, we can define the generating function $\hat{\Psi}(z) =
\sum_k z^k k P(k) \left[ \bar{k}_{nn}(k) -1 \right]$, that takes the
form
\begin{equation}
 \hat{\Psi}(z)  =\frac{\lambda
z}{T-1} \int_1^T dt e^{\bar{k}(t)(z-1)}\int_1^Tdt' r(t,t')
\bar{k}(t') \label{knn},
\end{equation}
from where it is possible to derive the large $k$ limit of the ANND.
We are primary interested in the case $\beta \geq 1/2$, that is, the
scale-free range of the model.  Focusing in $\beta>1/2$, the limit $T\to
\infty$ of Eq.~(\ref{knn}) is
\begin{equation}
 \hat{\Psi}(z)  =
\frac{\alpha^2 \lambda^2 z}{\beta(2\beta-1)} T^{2\beta-1} \int_0^1
\frac{1}{\tau^{1-\beta}} e^{(z-1)\bar{k}(\tau)} d\tau,
\end{equation}
where we have used the same change of variables as before. An
expansion of this equation around $z=1$ leads to
\begin{equation}
 \hat{\Psi}(z) \simeq
\frac{\alpha^2 \lambda^2 T^{2\beta-1}}{\beta^2(2\beta-1)} \left(
1+\frac{\alpha \lambda (1-z)}{\beta} \ln{\frac{\alpha \lambda
(1-z)}{\beta}}  \right).
\end{equation}
Applying again Tauberian theorems, we can write that, in the large $k$
limit,
\begin{equation}
k P(k) \left(\bar{k}_{nn}(k)-1 \right) \sim T^{2\beta-1}
\frac{1}{k^2},
\end{equation}
from where it is straightforward to derive the large $k$ limit of
the ANND:
\begin{equation}
\bar{k}_{nn}(k) \sim
N^{(3-\gamma)/(\gamma-1)}  k^{-(3-\gamma)}.
\end{equation}
The case $\beta=1/2$ is very similar to the previous one, except
for the type of divergence with the system size appearing as a
prefactor. In the large $T$ limit, we can write
\begin{equation}
\hat{\Psi}(z)  \sim 2 \alpha^2 \lambda^2 z \ln{T} \int_0^1
\frac{1}{\tau^{1/2}} e^{-2 \alpha \lambda (1-z)\tau^{-1/2}}d\tau.
\end{equation}
Using the same arguments we conclude that the right hand side of this
equation scales as $k^{-2}$ and, therefore, the ANND converges to a
constant value proportional to $\ln{N}$.

%\vspace*{0.5cm}

\section{Conclusions}
\label{sec:conclusions}

In this paper we have analyzed in detail a general class of complex
network models which are based on the existence of a hidden space, in
which vertices are located, and a connection probability that depends
on the hidden variable of each vertex.  The Markovian character at the
hidden level allows to calculate analytical expressions for the most
important structural properties, such as degree distribution, the ANND
function---quantifying two vertices correlations---and clustering
coefficient, as a measure of three vertices correlation. Our formalism
is valid for both sparse and non-sparse networks, extending the
applicability of our results to a wide range of complex networks.  At
this respect, one of the applications of our formalism is to provide
the analytical solution of a recently introduced model with intrinsic
fitness, which has recently attracted a great deal of interest as a
way to obtain scale-free networks without preferential attachment. Our
solution has been successfully contrasted with numerical simulations,
thus validating the accuracy of our formalism.

Another interesting result of our analysis is to provide a new
algorithm for generating correlated random networks with an {\it a
  priori} specified correlation structure. In this case, we also
calculate exact formulas for the relevant quantities. We have tested
the algorithm using a probe joint distribution $P(k,k')$, with very
encouraging results.

Perhaps the most striking result concerns the application of the
formalism to growing networks. Even though, in this case, the
network is out of equilibrium, it is possible to map it to a
specific kind of hidden variables model by the identification of
the injection time as the hidden variable. In order to check this
point we have applied the method to a general class of growing
networks, which, as a particular case, contains the
Barab{\'a}si-Albert model. Using our formalism, we have recovered
all the known results for this models, both for the degree
distribution and the correlation structure. It is remarkable that,
from a static approach, our formalism is able to derive correct
results for non-equilibrium evolving networks. Therefore, this
approach opens a novel and appealing way to study such systems.

\begin{acknowledgments}

  We thank G. Caldarelli, M. A. Mu{\~n}oz, and A. Vespignani for helpful
  comments and discussions.  This work has been partially supported by
  the European Commission - Fet Open project COSIN IST-2001-33555.
  R.P.-S. acknowledges financial support from the Ministerio de
  Ciencia y Tecnolog{\'\i}a (Spain).

\end{acknowledgments}

\end{document}